\documentclass[%
 reprint,
 amsmath,amssymb,
 aps,
]{revtex4-2}

\usepackage{graphicx}
\usepackage{dcolumn}
\usepackage{bm}
\usepackage{xcolor}
\usepackage{natbib}
\usepackage{tikz}
\usepackage{pgfplots}
\usepackage{xcolor}
\usepackage{hyperref}
\usepackage{gnuplottex}
\usepackage{amsmath}
\usepackage{lineno}

\newcommand{\hsn}[1]{\textcolor{black}{#1}}
\newcommand{\ab}[1]{\textcolor{black}{#1}}

\begin{document}

\preprint{APS/123-QED}

\title{Phases and correlations in active nematic granular systems}

\author{Abhishek Sharma}
\email{absharma12995@gmail.com}
\author{Harsh Soni}%
 
\affiliation{School of Physical Sciences (SPS), Indian Institute of Technology Mandi\\
 Kamand, Himachal Pradesh, India-175005
}%





\begin{abstract}
We investigate the statistical behavior of a system comprising fore-aft symmetric rods confined between two vertically vibrating plates using numerical simulations closely resembling the experimental setup studied by Narayan et al., Science 317, 105 (2007). Our focus lies in studying the phase transition of the system from an isotropic phase to a nematic phase as we increase either the rod length or the rod concentration. Our simulations confirm the presence of long-range ordering in the ordered phase. Furthermore, we identify a phase characterized by a periodic ordering profile that disrupts translation symmetry. We also provide a detailed analysis of the translational and rotational diffusive dynamics of the rods. Interestingly, the translational diffusivity of the rods is found to increase with the rod concentration. 
\begin{description}
\item[Key-words]
Active matter, granular materials, soft matter.
\end{description}
\end{abstract}

\maketitle


\section{\label{sec:level1}Introduction}

An active system is composed of individual components capable of converting energy from their environment into mechanical motion or other forms of activity~\cite{RevModPhys.85.1143}. These components span a wide range of length scales, from micron-sized particles such as synthetic microswimmers~\cite{liebchen2018synthetic,Bricard2013,bricard2015emergent,bricard2013emergence,sahu2020omnidirectional,brotto,hydro_thoery_band_prl_2014_Tailleur,morin2017distortion,kokot2018manipulation,zottl2016emergent}, microorganisms~\cite{bact_2004_prl_dombrowski_for_fig,PhysRevLett.93.098103,0034-4885-75-4-042601,lopez2015turning,sokolov2009reduction}, and biological cells~\cite{needleman2017active,doostmohammadi2018active,blow2014biphasic,saw2018biological,balasubramaniam2022active}, to granular active particles measuring a few millimeters~\cite{1,2,3,4,5,6,PhysRevLett.95.044101,kudrolli_pre_2003_granular_rods,PhysRevLett.100.058001,PhysRevLett.105.098001,polar_disk_long_sfot_matter}, and finally to macroscopic creatures like fish and birds~\cite{hemelrijk2012schools,lopez2012behavioural,parrish2002self,cavagna2014bird}.
Compared to conventional nonequilibrium systems, active systems exhibit many advanced features such as self-organization, pattern formation, and dynamic phase transitions, including more recently discovered phenomena like motility-induced phase separation~\cite{mips} and nonreciprocity~\cite{fruchart2020phase}.
Lab experiments on active systems include collective dynamics of microorganisms in vitro, molecular motor-based active entities like microtubules and actin~\cite{doostmohammadi2018active,Sanchez2012}, active colloids~\cite{Bricard2013,brotto,hydro_thoery_band_prl_2014_Tailleur}, and active granular \hsn{systems~\cite{1,2,3,4,5,6,PhysRevLett.96.028002,aranson2007swirling,galanis2010nematic,gonzalez2017clustering}.} The experimental setup of active granular systems, involving the collection of grains on vertically shaking plates~\cite{1,2,3,4,5,6}, offers the advantages of cost-effectiveness, easy assembly, and effortless executability. Despite that, much interesting physics of active systems have been discovered using these systems.

In this paper, our focus is on active apolar grains. Placing a rod-like particle on a vertically shaking plate enables it to tilt with respect to the plate, resulting in preferential motion along its axis rather than laterally. When the rod is fore-aft symmetric, its net motion vanishes, though it undergoes diffusion along its axis, resembling an active apolar particle~\cite{vijay_thesis}. Experimental studies with such rods have unveiled intriguing collective behaviours such as motile topological defects and long-lived large-number fluctuations~\cite{VijayScience2007}. However, these investigations were impeded by limitations like finite size and boundary effects, preventing a thorough understanding of the system's properties in the thermodynamic limit. While prior numerical endeavours employed simplified two-dimensional \hsn{models~\cite{Shi2013,chate2006simple}}, we simulate the system in three dimensions, accounting for rod collisions with vibrating plates, thus faithfully replicating the \hsn{actual experimental} system~\cite{1,2,3,4,5,6}. Our goal is to conduct a comprehensive analysis encompassing both statistical and dynamical properties of our system. \hsn{Our simulation model has already been quite successful in explaining the dynamics of a mixture of asymmetrically tapered motile rods and spherical beads~\cite{1,2,3,4,5,6}.}\\
Here are the key findings of the paper. We report a phase transition from  an isotropic phase to a nematic phase of the system as the concentration or length of the rods is increased. Saturating nematic order parameter correlation function with distance in the ordered phase indicates towards existence of the long-range ordering in the nematic phase. Moreover, at high rod concentrations, we observe a new phase in which the ordering profile of the rods breaks translation symmetry. The diffusion dynamics of the rods is also analysed by calculating the mean square displacement. The rods exhibit normal diffusion in both the disordered and ordered phases. Interestingly, the diffusion constant increases with rod concentration in the isotropic phase and along the direction of nematic ordering in the ordered phase. However, perpendicular to the alignment direction, the diffusion constant decreases with increasing rod concentration, a trend similar to equilibrium systems. Furthermore, we also present a study on the angular diffusion of the rods.

The remaining paper is organized as follows: Section~\ref{simdet} discusses simulation details. Section~\ref{results} presents the simulation results, beginning with the isotropic-nematic phase transition, followed by correlations in the system and then the diffusive properties of the system. Finally, in the last section, we conclude the paper with a discussion, a summary, and future possibilities of the work.
\section{Simulation Details}~\label{simdet}
\begin{figure}[h!]
\input{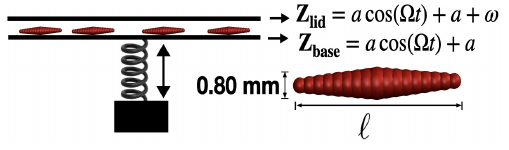}
\caption{(a) Schematic diagram of our system showing granular rods confined between two vertically vibrating plates; $Z_{\text{base}}$ and $Z_{\text{lid}}$ represent the $z$ coordinates of the base and lid plates, respectively.  (b) Apolar rod made of overlapping spherical beads.}
\label{fig01}
\end{figure}
Our system consists of many rods confined between two vertically oscillating plates (see Fig.~\ref{fig01}a). Each rod is rigid and geometrically apolar, with ends thinner than its middle part, matching the shape of the rods taken in~\cite{VijayScience2007}. For numerical simplicity, we construct the rod using an array of \hsn{equispaced} overlapping beads of varying diameters; the diameter is $0.8$ mm for the middle sphere and $0.4$ mm for both end spheres (see Fig.~\ref{fig01}b), unless otherwise stated. \hsn{As one goes from the center to either end of the rod, the diameter of the beads decreases linearly with respect to their distance from the central bead.} The length $\ell$ of the rod varies from $2.5$ mm to $4.5$ mm. The rods are placed between vertically shaking parallel hard walls with a spacing of $w=1.2$ mm, resulting in quasi-two-dimensional motion (Fig.~\ref{fig01}a); the $z$-coordinates of the walls at time $t$ are given by $Z_\text{base}(t)=a\cos(\Omega t)+a$ and \hsn{$Z_\text{lid}(t)=a\cos(\Omega t)+a+w$}, where $\Omega$ and $a$ are the angular frequency and the amplitude of the vibrations, respectively. The value of the dimensionless quantity $\Gamma = a\Omega^2/g$ is set to 7.0, where $g$ is the gravitational acceleration. \hsn{The parameter $\Gamma$ quantifies the activity of the system: when its value is zero, the system is inactive. The larger its value, the more active the system becomes.} In the $xy$ plane, we take the simulation box to be square-shaped, with periodic boundary conditions. The size of the simulation box in the $xy$ plane is chosen to be $L=83.73$ mm, unless otherwise stated. The rod-rod and rod-wall collisions are assumed to be instantaneous and are modelled using the impulse-based collision model~\cite{1,2,3,4,5}. \hsn{The mass of the rod is redundant for estimating the collision response with this model; only the inertia tensor, scaled by the mass, is required.} \hsn{In order to calculate the post-collision velocities using the impulse-based collision model~\cite{1,2,3,4,5}, we require the values of the coefficients of friction and restitution, $\mu$ and $e$. For different collisions, $\mu$ and $e$ are given in Table~\ref{tab:table1}. These values are chosen based on those selected in our recent work to reproduce the actual experimental results of a system of motile brass rods amidst a medium of spherical beads~\cite{kumar2014flocking}.} The rod's motion between two collisions is dictated by Newtonian rigid body dynamics \hsn{(see Appendix~\ref{simudet})}. We employ the time-driven particle dynamics algorithm for simulating our system. 
All simulation movies and snapshots are generated using VMD software~\cite{VMD}. \hsn{All the average quantities presented here are calculated in the steady state of the system. When the nematic order parameter stabilizes at a constant value, the system is considered to have reached a steady state. The typical simulation time for these systems depends on rod concentration and rod length, varying from 6,000 to 6,500 seconds, with a time step set to 4 microseconds.} \ab{The typical time to reach steady state varies from 400 seconds to 600 seconds, depending on rod concentration and length. Once the steady state is reached, the total simulation runs for an additional 6000 seconds.} 

In our study, we mainly vary two parameters: the rod length $\ell$ and the rod area fraction $\phi$ which is defined as
\begin{equation}
\phi=\dfrac{N A_\text{rod}}{L^2},
\end{equation}
where $N$ is the number of rods and $A_\text{rod}$ is the area of the projection of a single rod in the $xy$ plane when its axis is in the $xy$ plane. \hsn{As we vary \(\phi\) from 0.50 to 0.85, the number of particles \(N\) for our system ranges between 2,500 and 4,400 for \(\ell = 2.5\) mm and between 1,400 and 2,400 for \(\ell = 4.5\) mm. However, we also study the system with a box size of \(L = 167.46\) mm to see the impact of system size on the phase diagram, for which \(N\) goes up to 18,000 for \(\ell = 2.5\) mm and 9,500 for \(\ell = 4.5\) mm.
}
\begin{table}[hbt!]
\caption{\label{tab:table1}%
The coefficients of friction and restitution, $\mu$  and $e$, for different collisions}
\begin{ruledtabular}
\begin{tabular}{lcdr}
\textrm{Collision}&
\textrm{$\mu$}&&
\textrm{$e$}\\
\colrule
Particle-Particle & 0.05 && 0.3 \\
Particle-Base & 0.01 && 0.3 \\
Particle-Lid & 0.01 && 0.3 \\
\end{tabular}
\end{ruledtabular}
\end{table}
\section{Results}~\label{results}
\begin{figure}[h!]
\input{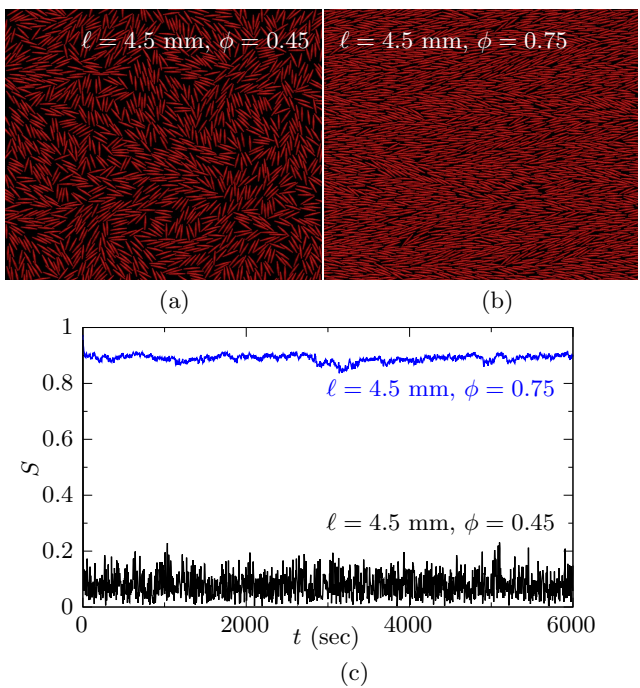}
\caption{(a) The steady-state configuration  of our system for $\ell=4.5$ mm and $\phi=0.45$; the system is disordered for this case. (b) The steady-state configuration for $\ell=4.5$ mm and $\phi=0.75$; the system is ordered for this case. (c) \hsn{Nematic order} parameter \hsn{$S(t)$} as the function of time $t$ for Figs. (a) and (b).} 
\label{fig02}
\end{figure}

\begin{figure}[h!]
\input{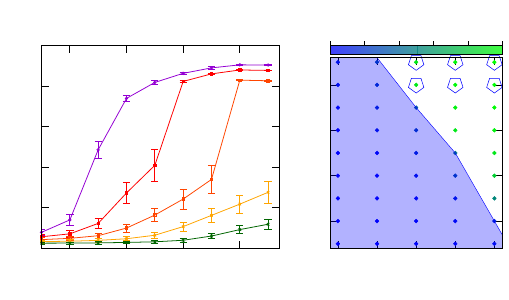}
\caption{(a) The nematic order parameter averaged over time in the steady state as the function of rod area fraction $\phi$ for different values of $\ell$. \hsn{The error bars indicate the standard deviation in the value of \(S(t)\) in the steady state.} (b) A phase diagram in the $\phi-\ell$ plane. \hsn{The blue pentagons depict the layered phase (see Fig.~\ref{fig07}a)}.
}
  \label{fig03}
\end{figure}

\begin{figure}
\input{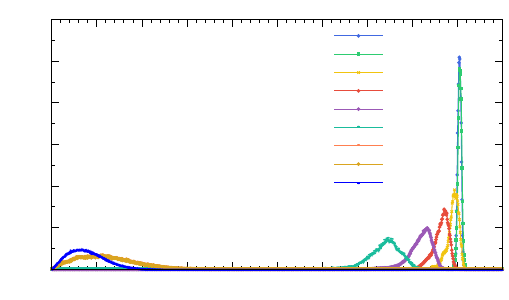}
    \caption{\hsn{The probability distribution \(P[S]\) of the nematic order parameter \(S(t)\) for \(\ell=4.5\) mm at different \(\phi\) values in the steady state.}}
    \label{figap01}
\end{figure}

At high $\phi$, it is easier to initialize the system with all the rods aligned in one direction; thus, we align all the rods in a single direction at $t=0$. \hsn{As the system's energy is not conserved at the microscopic level, the initial velocities and angular velocities of the particles are not crucial here; these values reach a steady-state distribution within a short time, on the order of the time period of the oscillations of the vibrating plates.} Subsequently, we conduct simulations for various $\ell$ and $\phi$ values, allowing the system to reach a steady state. All the analyses presented here are performed using the data obtained from the steady states of the system.
Figs.~\ref{fig02}(a) and (b) show that for $\ell=4.5$ mm, the rods are randomly oriented at $\phi=0.45$, whereas they are aligned  in a single direction (along the $x$-axis) at $\phi = 0.75$ \hsn{(see Supplementary Movies \hyperref[supmovie]{S1} and \hyperref[supmovie]{S2})}. This indicates a phase transition from an isotropic phase to a nematic phase as $\phi$ is increased.
To quantify the nematic order, we calculate the nematic order parameter of the system \hsn{$S(t)$}, which is given by
\begin{equation}
S(t)=\sqrt{\left\langle\cos 2\theta_i\right\rangle^2+\left\langle\sin 2\theta_i\right\rangle^2},
\end{equation}
where $\theta_i(t)$ is the angle of the $i$th rod with respect to the $x$-axis of the lab frame \hsn{at time $t$}, and the angular bracket $\left\langle\right\rangle$ denotes averaging over all the rods.
Fig.~\ref{fig02}(c) shows that \hsn{$S(t)$} fluctuates around zero for $\phi=0.45$ as the system evolves with time, indicative of the disorder in the system, as elucidated in Fig.~\ref{fig02}(a). Whereas \hsn{$S(t)$} remains around $0.9$ for $\phi=0.75$, reflecting the nematic ordering of the rods, as illustrated in Fig.~\ref{fig02}(b). Note that the value of \hsn{$S(t)$} is 1 at $t=0$ for both cases as the rods in our initial configuration are perfectly aligned in one direction, as discussed earlier.
The plots of the time-average $\left\langle S\right\rangle_t$ of the nematic order parameter \hsn{$S(t)$} as the function of $\phi$ for various values of $\ell$ are shown in Fig.~\ref{fig03}(a): for the values of $\ell$ above 3.5 mm, $\left\langle S\right\rangle_t$ continuously increases from 0 to around 1 which hints towards the second order phase transition from an isotropic phase to a nematic phase. 
For  $\ell$ below 3.5 mm, the system always remains in the isotropic phase \hsn{(see Supplementary Movies \hyperref[supmovie]{S3} and \hyperref[supmovie]{S4})}.
The reason small rods fail to align can be understood in terms of Onsager's theory of volume exclusion for hard rods~\cite{onsager1949}. Since small rods have less excluded volume, they can easily rotate and therefore cannot remain ordered.
Using the values of $\langle S \rangle_t$, we construct a phase diagram in the $\phi - \ell$ plane, as illustrated in Figure ~\ref{fig03}(b). \hsn{For infinitely large systems, a system with \(\langle S \rangle_t > 0\) would be called an ordered system. However, considering the finite system size, we instead use the criterion that the system is said to be ordered when \(\langle S \rangle_t > 0.5\); otherwise, it is called disordered. For our system, this criterion aligns with the expectation that the nematic order parameter correlation function should decay to zero in the disordered phase and saturate to a nonzero value in the ordered phase (discussed later).}
\hsn{It becomes evident from the phase diagram that, for a given rod length $\ell$, the system exhibits nematic order above a threshold value $\phi_c$. This threshold $\phi_c$ decreases with increasing $\ell$.} Thus, within the region characterized by large $\phi$ and large $\ell$, the system predominantly adopts a nematic ordered state. Conversely, in the region with small $\phi$ and small $\ell$, the system tends to remain in an isotropic state. 
We acknowledge that the phase boundary is influenced by the system size. Due to the high computational cost of these simulations, we cannot perform all analyses with a larger system size. However, we have obtained a phase diagram for a larger system with $L = 167.46$ mm (see Fig.~\ref{phasdialarge} in the Appendix). The critical value of $\phi$ for a given $\ell$ increases with the system size $L$.
\hsn{The nematic phase transition in our system can be further understood by examining the probability distribution \(P[S]\) of the nematic order parameter \(S( t)\) in the steady state as \(\phi\) decreases for a given \(\ell\).} 
\hsn{As displayed in Fig.~\ref{figap01}, for $\ell=4.5$ mm the} distribution is always unimodal, suggesting the presence of only one predominant phase in our system. However, there is a noticeable shift in the peak position of the distribution from around $S \approx 1$ to values closer to $S = 0$ as the transition occurs  This unimodal characteristic of the distribution functions suggests the absence of any metastable phase on the opposite side of the transition, which is a hallmark feature of continuous transitions. Additionally, as the system approaches the phase transition point, the distributions become broader, reflecting heightened fluctuations within the system.

These observations are further reinforced by examining the nematic order parameter correlation function $G_2(r)$ which is defined as follows: 
\begin{equation}
G_2(r)=\left\langle\cos\left[2(\theta_i -\theta_j)\right]\right\rangle_r,
\end{equation}
where $\left\langle\right\rangle_r$ represents \hsn{a steady-state} average over all pairs $(i,j)$ of rods separated by the distance $r$.
\begin{figure}[h!]
\input{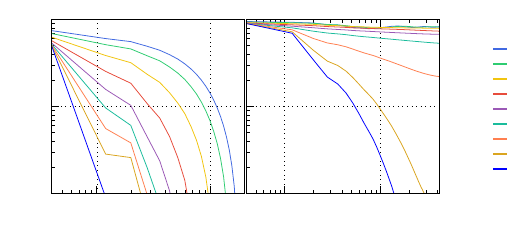}
\caption{The nematic order parameter correlation function $G_2(r)$ on log-log scale for various values of $\phi$ at (a) $\ell=$2.5 mm and (b) $\ell=4.5$ mm.}
\label{fig04}
\end{figure}
Fig.~\ref{fig04} illustrates the dependence of $G_2(r)$ on $r$.
For $\ell=2.5$ mm, the system remains disordered, resulting in $G_2(r)$ decaying with distance to zero for all values of $\phi$ (see Fig. \ref{fig04}a). However, for $\ell=4.5$ mm, $G_2(r)$ saturates to a constant for large values of $\phi$, indicating the presence of global order in the system (see Fig. \ref{fig04}b).  For this case, for small $\phi$, $G_2(r)$ diminishes with $r$ as the system is disordered. To capture the anisotropy in spatial correlations of the nematic order parameter, we calculate the nematic order parameter correlation function $G_2(\mathbf{r})$ as a function of the relative position vector $\mathbf{r}$ and then construct its heat map, as depicted in Fig.~\ref{fig05}. In the disordered state, $G_2(\mathbf{r})$ remains isotropic, whereas in the ordered state, it declines faster in the direction normal to the average alignment [see Figs.~\ref{fig05}(a) \& (c)]. These trends are more evident from Figs.~\ref{fig05}(b) \& (d), which illustrate the dependence of $G_2(\mathbf{r})$ on distance $r$ along and normal to the direction of alignment. In terms of Frank elasticity, this observation suggests that the Frank elastic constant for splay deformations is larger than that for bending deformations, which is unsurprising given the large aspect ratio of our rods~\cite{de1993physics}.
\begin{figure}[h!]
\input{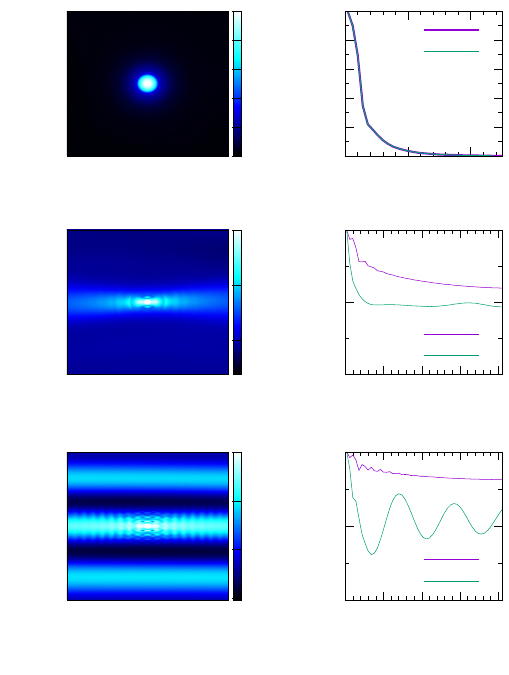}
\caption{Anisotropy in the nematic order parameter correlation function $G_2(\mathbf{r})$ for $\ell =4.5$ mm for different $\phi$. (a) Heat map and (b) $G_2(\mathbf{r})$ vs $r$  along $x$ and $y$ for $\phi = 0.45$. (c) Heat map and (d) $G_2(\mathbf{r})$ vs $r$ parallel and perpendicular to alignment for $\phi =0.75$ . (e) Heat map and (f) $G_2(\mathbf{r})$ vs $r$ along and normal to the alignment direction for $\phi =0.85$.}
\label{fig05}
\end{figure}

\hsn{In the ordered phase, for high area fractions ($\phi \geq 0.80$, see Fig.~\ref{fig03}b),} a novel phase of the system emerges wherein the orientation of the rods exhibits a periodic structure, as depicted in Fig.~\ref{fig07}a (see Supplementary Movie \hyperref[supmovie]{S5}). This periodicity is reflected in the heat map of $G_2(\mathbf{r})$, showing oscillations normal to the direction of the ordering [see Fig.~\ref{fig05}e]; the plot of $G_2(\mathbf{r})$ vs $r$ normal to the direction of the nematic alignment is displayed in Fig.~\ref{fig05}f.
\begin{figure}[h!]
\input{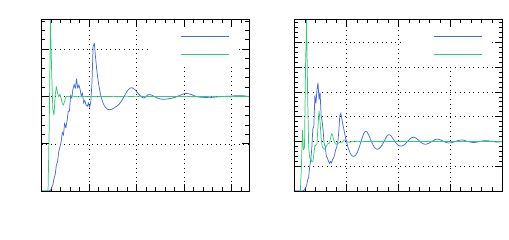}
\caption{The pair distribution function $g(r)$ in the nematically ordered phase for (a) $\phi=0.75$ and (b) $\phi=0.85$.}
\label{fig06}
\end{figure}
In this phase, the distribution of the orientation angle $\theta$ of the rods relative to the direction of the net nematic alignment exhibits two peaks slightly offset from zero, suggesting that the rods tend to orient themselves at angles $\theta=\pm\theta_m$ with respect to the alignment direction, where $\theta_m\simeq 10$ degrees [see Fig.~\ref{fig07}b]. In this phase, the system exhibits a layered structure reminiscent of a smectic phase; however, alternate layers feature different average rod angles, $\theta=\theta_m$ and $\theta=-\theta_m$. \hsn{Fig.~\ref{fig07}b also suggests that \(\theta_m\) increases with \(\phi\).
The probability distribution of \(\theta\) for different values of the radius \(r_t\) of the end spheres of the rods can be seen in Fig.~\ref{fig07}c. As shown in Figs.~\ref{fig07}d \& e, \(\theta_m\) decreases with rod length \(\ell\) as well as with \(r_t\).
}
Thus, the layered structure can be attributed to the tapered shape of the apolar rod.
We further plot the pair distribution function $g(r)$ for our system in Fig.~\ref{fig06}. Along the alignment direction, $g(r)$ quickly flattens with $r$ compared to the perpendicular direction, indicating that rods exhibit more local spatial ordering in the direction of alignment than perpendicular to it. 
\begin{figure}[h!]
\input{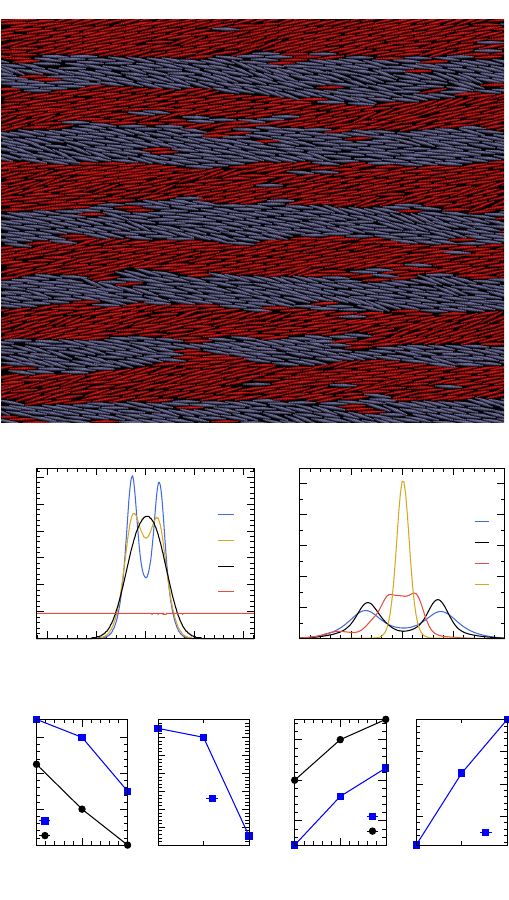}
\caption{(a) Steady-state configuration of the system for $\ell=4.5$ mm and $\phi=0.85$. \hsn{The rods with orientation angle \(\theta > 0\) are shown in red, while the rods with \(\theta < 0\) are shown in magenta, here system is aligned along $x$ direction.} (b) Distribution function $P(\theta)$ of the orientation angle $\theta$  of the rods with respect to the axis of the net nematic alignment of the system for different $\phi$ at $\ell=4.5$ mm. For \hsn{$\phi=0.45$}, $\theta$ is with respect to the $x$ axis of the lab frame. (c)  $P(\theta)$ vs $\theta$ for different values of the radius $r_t$ of the ends of the rods. \hsn{(d) \(\theta_m\) vs $\ell$ for $\phi=0.80$ and $\phi=0.85$. (e) \(\theta_m\) vs \(r_t\) for $\phi=0.85$. (f) Wavelength \(\lambda\) of the layered pattern vs \(\ell\) for $\phi=0.80$ and $\phi=0.85$. (e) \(\lambda\) vs \(r_t\) for $\phi=0.85$.}}
\label{fig07}
\end{figure}
\hsn{We now calculate the wavelength \(\lambda\) of the layered patterns by analyzing \(G_2(\mathbf{r})\) normal to the alignment axis. We observe that \(\lambda\) increases as \(\phi\) decreases from 0.85 to 0.80. Additionally, it raises with the rod length as well as with the radius \(r_t\) of the end beads (see Figs.~\ref{fig07}f \& g).  Our simulations with a larger system size \(L = 167.46\) mm reveal that \(\lambda\) increases while \(\theta_m\) decreases with system size.} 

\hsn{The layer formation is also related to the dynamics of the rods in the third direction. This is confirmed by the observation that, for $\phi=0.80$, when the gap between the plates \(w\) is reduced to 0.9 mm, the system forms a crystalline order with some dislocations. Whereas, when the gap is widened to 1.4 mm, again the layers disappear, though the system remains in an ordered phase (see Fig.~\ref{gapif} of the appendix).}

We further examine the diffusion dynamics of the rods in our system. In the nematic phase, the system is anisotropic, so it is more appropriate to calculate the components of the mean square displacement (MSD) of the rod along the direction of the average alignment and normal to it, denoted as $\langle\Delta r^2_\parallel(t)\rangle$ and $\langle\Delta r^2_\perp(t)\rangle$, respectively. Here, \hsn{the} angular brackets denote the average over all rods presented in the system \hsn{and over the initial time of the trajectory.}
\begin{figure}[h!]
\input{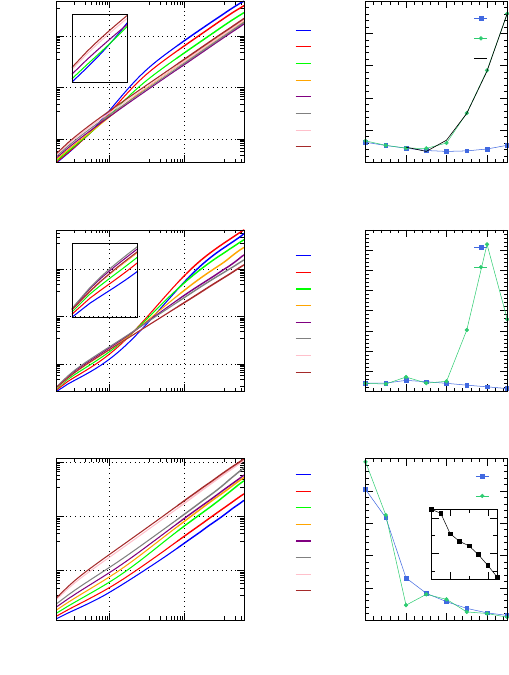}
\caption{Log-log plot of the Mean Square Displacement (MSD) of the rods as a function of time $t$ for varying area fractions $\phi$: (a) for the isotropic phase with $\ell=2.5$ mm; (c) for the nematic phase along the nematic alignment direction with rod length $\ell=4.5$ mm; (e) for the nematic phase normal to the nematic alignment direction with rod length $\ell=4.5$ mm. Panels (b), (d), and (f) show $D$ vs $\phi$ for $t \to 0$ and $t \to \infty$ corresponding to (a), (c), and (e), respectively. The inset of (f) shows the MSD scaling exponent $\alpha$ vs $\phi$ in the $t \to 0$ limit. In (b), the black solid line shows the fit for $t \to \infty$ beyond $\phi = 0.6$. The fit is given by $D - D_0 = C(\phi - \phi_0)^2$, where $D_0 = 1.7$ mm$^2$s$^{-1}$, $C = 480$ mm$^2$s$^{-1}$, and $\phi_0 = 0.64$.}
\label{fig09}
\end{figure}
Fig.~\ref{fig09}a illustrates the behavior of the mean square displacement (MSD) of rods for the isotropic state, considering a rod length of $\ell = 2.5$ mm, across varying area fractions $\phi$. At small as well as large time scales, the rods diffuse normally. The diffusion constant in the large $t$ limit increases with $\phi$ (see Fig.~\ref{fig09}b). 
This is contrary to equilibrium systems, where the diffusion constant typically decreases with particle number density. 

A similar trend is seen along the direction \hsn{$\mathbf{N}$} of the net alignment when the system is ordered, as shown in Figs.~\ref{fig09}c \& d, except that, beyond $\phi=0.80$, the diffusion coefficient declines due to the almost frozen dynamics of the rods. On the contrary, we observe that the diffusion constant of the rods normal to \hsn{$\mathbf{N}$} declines with increasing $\phi$ (see Figs.~\ref{fig09}e \& f). Moreover, a visual comparison between Fig.~\ref{fig09}d and Fig.~\ref{fig09}f indicates that the rods diffuse more along \hsn{\(\mathbf{N}\)} than perpendicular to it. \hsn{This is because the rods predominantly move along \(\mathbf{N}\), and for motion perpendicular to \(\mathbf{N}\), the effective collision cross-section is much larger—approximately of the order of the rod's length—compared to the cross-section for motion along the alignment axis, which is of the order of the rod's diameter.}
Another interesting observation is that, at small time scales, the rods become sub-diffusive in the normal direction at high $\phi$ values. The inset of Fig.~\ref{fig09}f shows the MSD scaling exponent $\alpha$, defined by the relation MSD$= \hsn{2}Dt^\alpha$, as a function of $\phi$. In contrast to the prediction for polar rods by~\cite{tonertu_pre_1998}, we do not observe superdiffusive behavior normal to the ordering direction.

\begin{figure}[h!]
    \input{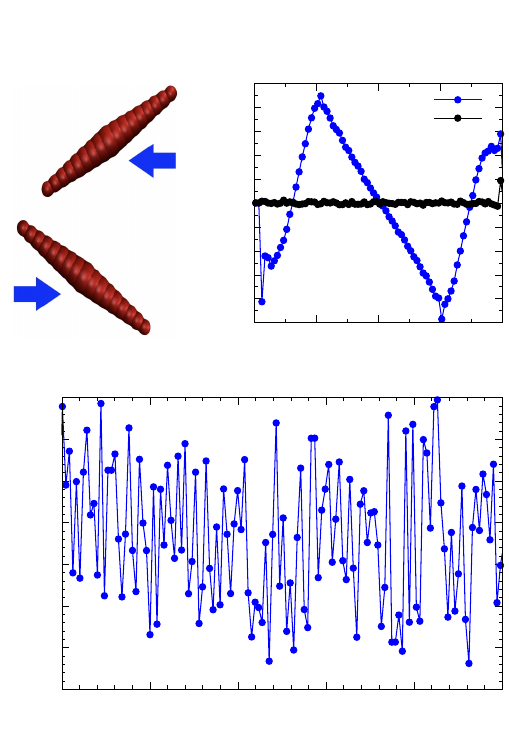}
    \caption{\hsn{(a) Schematic diagram illustrating the connection between the tilt of the rod with respect to the horizontal plane and its average direction of motion; the blue arrow indicates the rod's average velocity direction for \(|n_z| < 0.3\), which reverses for \(|n_z| > 0.3\). (b) The average velocity components \(v^\text{n}\) and \(v^\text{t}\) of the rod along and perpendicular to the rod axis as a function of the \(z\) component of the rod's orientation \(n_z\) in the steady state. (c) \(n_z\) vs time for a single rod in the steady state. Here $\phi=0.6$ and $\ell=2.5$ mm.}}
    \label{vvsnz}
\end{figure}

\hsn{Before understanding the dependence of the diffusion constant on \(\phi\), it is essential to first grasp how the rods can move in the \(xy\) plane. The rods are fore-aft symmetric but have the ability to tilt with respect to the \(xy\) plane, given the finite gap between the plates. The tilt of the rod spontaneously breaks fore-aft symmetry, causing the rod to begin moving in one direction along its axis due to the friction at the contact points during rod-plate collisions (see Fig.~\ref{vvsnz}a). To quantify this observation, we calculate the average component \(v^\text{n}_i\) of the rod's velocity along its orientation \((\cos\theta_i, \sin\theta_i)\) in the \(xy\) plane as a function of the \(z\) component \(n_z\) of the orientation vector in three dimensions:
\begin{equation}
v^\text{n}(n_z) = \left\langle v^\text{n}_i \right\rangle_{n_z},
\end{equation}
where \(\left\langle  \right\rangle_{n_z}\) denotes the steady-state average over all rods with a \(z\) component of the orientation vector equal to \(n_z\).
As shown in Fig.~\ref{vvsnz}b, \(v^\text{n}\) exhibits a strong dependence on \(n_z\); it is zero when \(n_z\) is zero and reaches its maximum magnitude around \(n_z = \pm0.2\). Since the sign of \(v^\text{n}\) is opposite to the sign of \(n_z\), except for a small range \(0.4 > |n_z| > 0.3\), the rods, on average, tend to move in the direction from the upper end to the lower end of the tilted rod, as illustrated in Fig.~\ref{vvsnz}a. The value of \(n_z\) for a rod varies with time (see Fig.~\ref{vvsnz}c). Therefore, the rod does not persistently move in one direction along its axis, unlike active Brownian particles. Instead, the rod keeps switching direction back and forth. As expected by symmetry, the average velocity component \(v^\text{t}\)of the rod normal to its orientation, calculated in the same manner, remains zero for all values of \(n_z\) (see Fig.~\ref{vvsnz}b).}

\begin{figure}[h!]
    \input{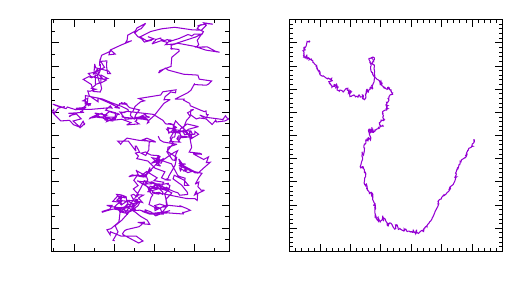}
    \caption{\hsn{Trajectory of a single rod in the steady state for a time interval of 30 seconds for a rod with \(\ell = 2.5\) mm: (a) \(\phi = 0.65\), (b) \(\phi = 0.85\).}}
    \label{trajects}
\end{figure}

\begin{figure}[h!]
\input{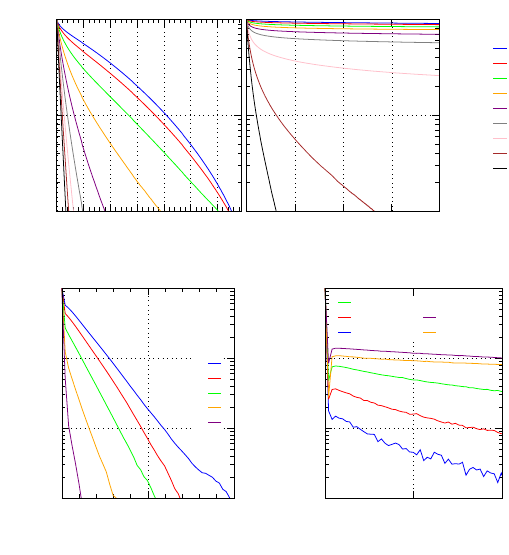}
\caption{The orientational autocorrelation function of the rods $C_2(t)$ for varying values of $\phi$ at rod lengths of (a) $\ell=2.5$ mm and (b) $\ell=4.5$ mm. \hsn{For \(\ell = 2.5\) mm, the system does not achieve ordering even at high \(\phi\), allowing the rods to explore all possible orientations in the \(xy\) plane. Thus, \(C_2(t)\) decays to zero. In contrast, for \(\ell = 4.5\) mm, \(C_2(t)\) stabilizes to a non-zero value close to one at high \(\phi\), as the rods' orientations are confined around the direction of the global alignment. (c) The velocity autocorrelation function \(C_v(t)\) for \(\ell = 2.5\) mm. (d) The velocity autocorrelation function \(C^\parallel_v(t)\) for the component along the alignment direction \(\mathbf{N}\) for \(\ell = 4.5\) mm.}}
\label{fig08}
\end{figure}

\hsn{We now provide a justification for the anomalous \(\phi\) dependence of the diffusion constant in the disordered phase. The trajectories of a rod at \(\phi = 0.65\) and \(\phi = 0.85\), illustrated in Figs.~\ref{trajects}a and \ref{trajects}b, reveal that as \(\phi\) increases, the rod tends to spend more time moving in one direction before changing direction (see Supplementary Movie \hyperref[supmovie]{S6}).
This behavior can be characterized by analyzing the orientation autocorrelation function \(C_2(t)\) and the velocity-velocity autocorrelation function \(C_v(t)\) of the rods, defined as follows:
\begin{eqnarray}
C_2(t) &=& \left\langle \cos\left\{2[\theta_i(t_0 + t) - \theta_i(t_0)]\right\} \right\rangle_i, \\
C_v(t) &=& \left\langle \mathbf{v}_i(t_0) \cdot \mathbf{v}_i(t_0 + t) \right\rangle_i.
\end{eqnarray}
Here, the angular brackets denote an average over all rods and time \(t_0\), with \(\theta_i(t)\) and \(\mathbf{v}_i(t)\) representing the orientation angle and velocity of the \(i\)-th rod at time \(t\). To filter out high-frequency noise in the data, \(\mathbf{v}_i(t)\) is computed by averaging the actual velocity over a period of 400 milliseconds.
As illustrated in Fig.~\ref{fig08}a, the decay rate of \(C_2(t)\) decreases with \(\phi\). Due to the reduced rotational noise, the velocity of the rods becomes more correlated over time. This is shown in Fig.~\ref{fig08}d, where \(C_v(t)\) declines more slowly with \(t\) at higher \(\phi\). Therefore, the enhanced local ordering with rising \(\phi\) results in a greater persistence length of the rods, which in turn enhances diffusion.
}

\hsn{In the ordered phase, the orientation autocorrelation function \(C_2(t)\) reaches a constant value, which increases with \(\phi\) (see Fig.~\ref{fig08}b). Consequently, at higher \(\phi\), the rods exhibit more persistent movement along the nematic alignment direction \(\mathbf{N}\). As a result, the velocity autocorrelation function for the component along \(\mathbf{N}\), \(C^\parallel_v(t) = \left\langle (\mathbf{v}_i(t_0) \cdot \mathbf{N}) (\mathbf{v}_i(t_0 + t) \cdot \mathbf{N}) \right\rangle_i\), also grows with $\phi$ up to $\phi=0.80$ (see Fig.~\ref{fig08}d). This leads to a higher diffusion constant along \(\mathbf{N}\) with increasing \(\phi\). As the rod's movements become progressively aligned with the nematic direction \(\mathbf{N}\) for higher \(\phi\), the diffusion constant perpendicular to \(\mathbf{N}\) declines.}



\begin{figure}[h!]
    \input{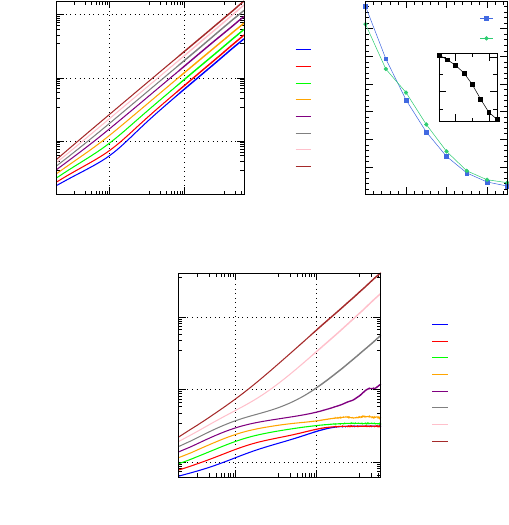}
 \caption{(a) Mean square angular displacement (MSAD) of the rods $\langle \Delta \theta^2 \rangle$ vs time $t$ for $\ell = 2.5$ mm. (b) The rotational diffusion $D_r$ vs $\phi$ for $\ell = 2.5$ mm in the $t \to 0$ and $t \to \infty$ limits; the inset shows the MSAD scaling exponent $\alpha_r$ as a function of $\phi$. (c) $\langle \Delta \theta^2 \rangle$ vs $t$ for varying $\phi$ at $\ell = 4.5$ mm.}
    \label{fig10}
\end{figure}

Rotational diffusion is examined by measuring the mean square angular displacement $\langle \Delta \theta^2 \rangle$ of the rods (see Fig.~\ref{fig10}). In the isotropic phase, the rods exhibit sub-diffusive rotation dynamics in the small $t$ limit, and the corresponding scaling exponent $\alpha_r$, defined as $\langle \Delta \theta^2 \rangle = 2D_r t^{\alpha_r}$, decreases with increasing $\phi$. In the large $t$ limit, normal rotational diffusion is observed. In contrast to translational diffusion, rotational diffusion decreases with $\phi$. In the nematic phase, $\langle \Delta \theta^2 \rangle$ saturates to a constant value, as shown for the $\ell = 4.5$ mm case in Fig.~\ref{fig10}c. For the same value of $\ell$, at lower $\phi$, $\langle \Delta \theta^2 \rangle$ shows a linear increase with $t$, since the system is disordered.

\section{Discussion and conclusion}
The phase transition in active nematics has been widely studied, encompassing theoretical, experimental, and numerical approaches~\cite{marchetti2013hydrodynamics,blow2014biphasic,Sanchez2012,doostmohammadi2018active,saw2018biological,balasubramaniam2022active,chate2006simple,tan20212d,D3SM01224G,D2SM00414C}.  Narayan et al.~\cite{VijayScience2007,narayan2006nonequilibrium} focused on experimental studies of elongated granular rods with specific rod lengths and limited area fractions. However, a detailed investigation of the nematic granular system was lacking until now. While we present a numerical study, our system closely mimics the actual experimental setup of the nematic granular system used by~\cite{VijayScience2007}. Our study provides a comprehensive examination across various rod lengths and encompasses a wider range of area fractions, thereby offering a detailed and extensive understanding of the isotropic-nematic phase transition in this system. Our primary motivation to explore this system stems from our goal to investigate the dynamics of motile particles in a nematic medium using vibrated granular systems.
\begin{figure}[h!]
    \centering
    \includegraphics[width=0.48\textwidth,height=0.48\textwidth]{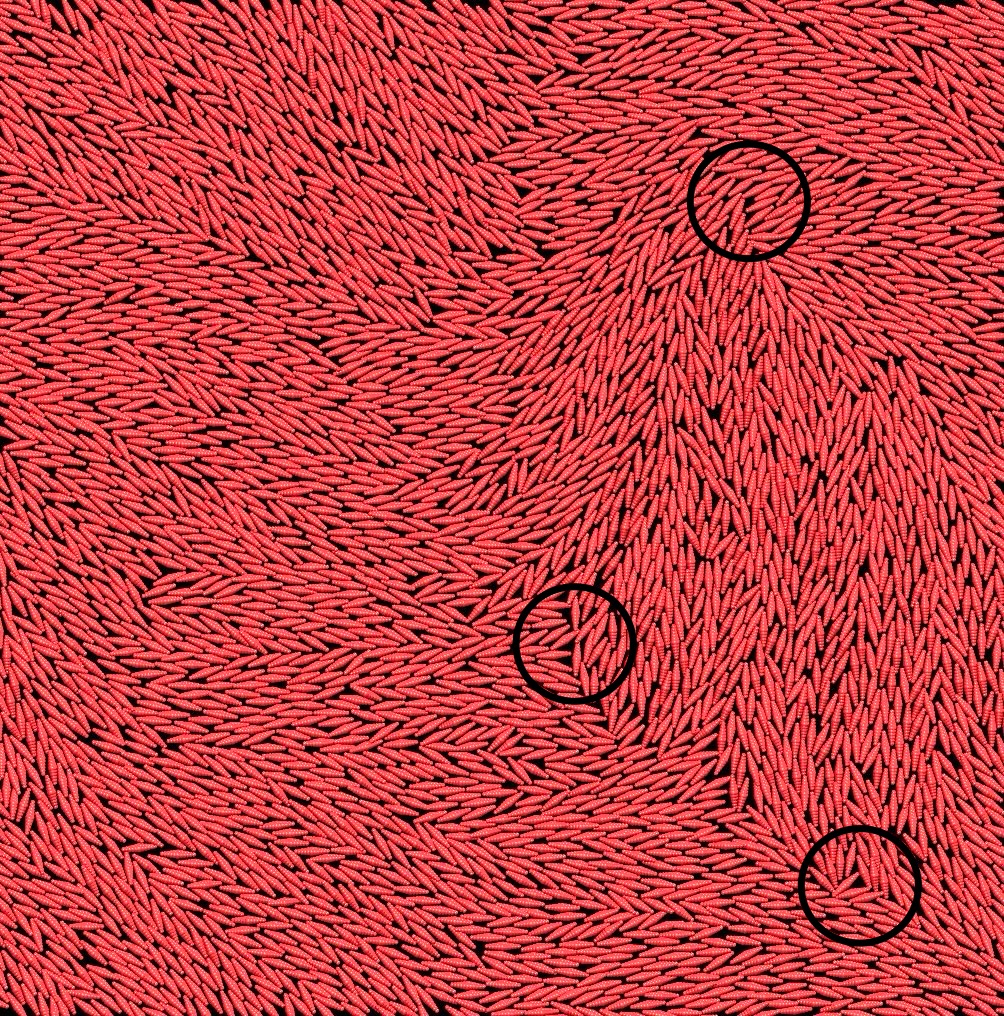}
    \caption{Topological defects (encircled in black) in the steady state for the case of $\ell=3.0$ mm and $\phi=0.85$.}
    \label{defectfig}
\end{figure}
This necessitates an understanding of the fundamental properties of granular nematic systems. The work presented here will lay the foundation for addressing such problems. 

The presented study lacks an understanding of the dynamics of defects in our system, which we aspire to explore in the future. One measure in this direction would be to investigate the mobility of topological defects and the consequent melting of nematic ordering. As seen in Fig.~\ref{defectfig} for $\ell=0.3$ mm and $\phi=0.85$, the system appears locally ordered but fails to achieve global nematic order due to the presence of defects \hsn{(also see Supplementary Movie \hyperref[supmovie]{S7})}. \hsn{In addition, we also plan to explore the impact of the activity on the system's dynamics by varying the value of the parameter $\Gamma$.}\\
To summarize, we conducted numerical studies to emulate the experimental setup of vibrated active nematic rods~\cite{1,2,3,4,5}. Our findings reveal an isotropic-nematic transition in our system as the packing area fraction increases. Additionally, we observed that the critical area fraction for the transition depends on the shape anisotropy; longer rods exhibit a lower critical area fraction, while smaller rods have a higher critical area fraction. We also investigated the spatial and temporal orientation correlations in our system, uncovering a novel phase of elongated granular rods that breaks translational symmetry, attributed to the tapered nature of the rods. Furthermore, we provided an elaborate study on the diffusive dynamics of the nematic rods, revealing enhanced diffusion with increasing rod concentration.

\section{Conflicts of Interest}
There are no conflicts to declare.

\section{Acknowledgements}
AS acknowledges Param Himalaya (NSM) for providing the computational resources and Prime Minister's Research Fellows (PMRF) for the financial support. HS acknowledges SERB for the SRG grant (no. SRG/2022/000061-G). 

\section{Appendix}
\subsection{\hsn{Equations of motion for a single rod}}\label{simudet}
As discussed in the main text, the rods interact with each other and the wall instantaneously through hard collisions in our system. When a collision occurs, the particles involved in the collision experience changes in both their angular and linear velocities. The collisions are modeled using the impulse-based collision model~\cite{1,2,3,4,5}. Here, we discuss the equations of motion for the rod between two consecutive collisions.

Apart from the collisions, rods are subject to no force other than gravity; hence, the rod performs free rigid body dynamics under gravity between two consecutive collisions.
The translational motion of the rod is straightforward: the equations of motion for the position \(\mathbf{r}_i(t)\) and the velocity \(\mathbf{v}_i(t)\) of the center of mass of the \(i\)-th rod are given by:
\begin{eqnarray}
    \mathbf{v}_i(t) &=& \mathbf{v}_i(t_0) - \hat{\mathbf{z}}g (t - t_0), \\
    \mathbf{r}_i(t) &=& \mathbf{r}_i(t_0) + (t - t_0)\mathbf{v}_i(t_0) - \hat{\mathbf{z}}g \frac{(t - t_0)^2}{2},
\end{eqnarray}
where \(t_0\) is the time of the last collision of the rod before time \(t\) and \(g\) is the gravitational acceleration.

Next, we consider the rotational motion of the rod. The equation of motion for the unit vector \(\mathbf{n}_i(t)\) along the axis of the rod is given by:
\begin{eqnarray}\label{neq}
    \dot{\mathbf{n}}_i(t) = \bm{\omega}_i(t) \times \mathbf{n}_i(t),
\end{eqnarray}
where \(\bm{\omega}_i(t)\) is the angular velocity of the rod. Since the rod is subject to no torque, its angular momentum \(\mathbf{J} = \bm{I}^\text{L}(t) \cdot \bm{\omega}_i(t)\) remains conserved, where \(\bm{I}^\text{L}\) is the inertia tensor of the rod in the lab frame. Therefore,
\(\dot{\bm{I}}^\text{L}(t) \cdot \bm{\omega}_i(t) + \bm{I}^\text{L}(t) \cdot \dot{\bm{\omega}}_i(t) = 0\). This gives us the following equation of motion for \(\bm{\omega}_i(t)\):
\begin{eqnarray}\label{omeq}
    \dot{\bm{\omega}}_i(t) = -\bm{I}^\text{L}(t)^{-1} \cdot \dot{\bm{I}}^\text{L}(t) \cdot \bm{\omega}_i(t).
\end{eqnarray}
The inertia tensor \(\bm{I}^\text{L}(t)\) in the lab frame also evolves with time as the rod rotates and is given by:
\begin{eqnarray}
    \bm{I}^\text{L}(t) = \bm{R}(t) \cdot \bm{I}^\text{B} \cdot {\bm{R}(t)}^\intercal,
\end{eqnarray}
where \(\bm{I}^\text{B}\) is the inertia tensor of the rod in the body frame and \(\bm{R}(t)\) is the transformation matrix from the body frame to the lab frame.
Since our rod is axisymmetric, it is convenient to choose the body frame in which the axis of the rod is along the $z$ direction. In that case, \({I}^\text{B}_{ij} = I_\perp \delta_{ij} + (I_\parallel - I_\perp) \delta_{i3} \delta_{j3}\), where \(I_\perp\) and \(I_\parallel\) are the moments of inertia of the rod with respect to the axes normal to the rod's axis passing through the center of mass and with respect to the rod's axis, respectively. In that case, 
\begin{equation}
\bm{I}^\text{L}(t) = I_\perp \mathbf{I}_3 + (I_\parallel - I_\perp) \mathbf{n}_i(t) \mathbf{n}_i(t),    
\end{equation}
 where \(\mathbf{I}_3\) is the \(3 \times 3\) identity matrix. Using the above expression, Eqs.~\eqref{neq} and~\eqref{omeq} are simulated numerically to obtain the trajectory of \(\bm{\omega}_i(t)\) and \(\mathbf{n}_i(t)\) between two successive collisions of the \(i\)-th rod.

\subsection{Phase diagram for a bigger system size}
\begin{figure}[h]
   \input{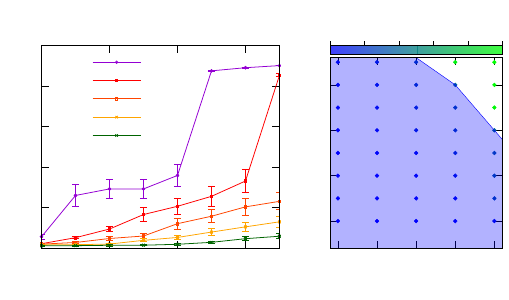}
    \caption{\hsn{Phase transition for \(L = 167.46\) mm. (a) The time-averaged nematic order parameter \(\left\langle S \right\rangle_t\) as a function of \(\phi\). (b) Phase diagram in the \(\phi-\ell\) plane.}}
    \label{phasdialarge}
\end{figure}
\subsection{Influence of the gap between shaking plates on the formation of the layered phase}

\begin{figure}[h!]
   \input{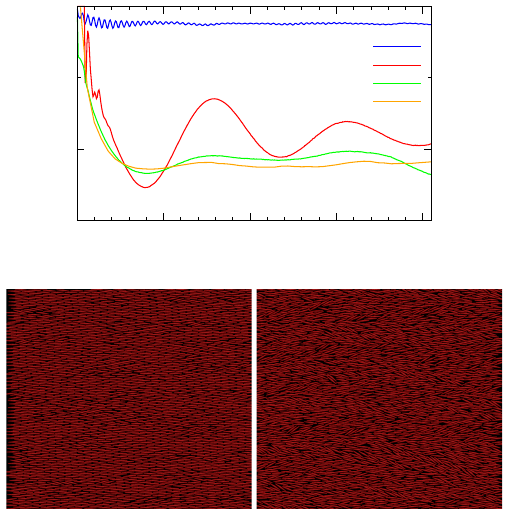}
    \caption{\hsn{(a) Nematic order parameter correlation function \(G_2(\mathbf{r})\) vs distance \(r\) normal to the alignment direction for varying gap \(w\) between the vibrating plates for \(\ell = 4.5\) mm and \(\phi = 0.80\). The oscillations in \(G_2(\mathbf{r})\) disappear as \(w\) is increased or decreased, indicating the dissolution of the layered phase. Screenshots of the system for \(\ell = 4.5\) mm and \(\phi = 0.80\); (b) for \(w = 0.9\) mm, and (c) for \(w = 1.4\) mm.}}
    \label{gapif}
\end{figure}

\newpage
\bibliographystyle{apsrev4-2}
\bibliography{paper01}
\newpage
\begin{widetext}
\section*{Information of Supplementary Data}
\subsection*{Supplementary  Movies}
\label{supmovie}
\begin{itemize}
    \item S1: This is a simulation movie for large rod length with $\ell=4.5$ mm, $\phi=0.45$ when system remains in isotropic state. \href{https://drive.google.com/file/d/17z0p6TIduAKOSjye7AAjSrG9tBhGpnOE/view?usp=share_link}{\color{blue}Movie here} 
    \item S2: This is a simulation movie for $\ell=4.5$ mm, $\phi=0.75$ when system gets ordered along $x$ direction.\href{https://drive.google.com/file/d/1xF4K9Emo_BlhSsBpxXSbbxBPRUoIbsnc/view?usp=share_link}{\color{blue}Movie here}
    \item S3: This is a simulation movie for small rod length with $\ell=2.5$ mm, $\phi=0.60$ when system remains in isotropic state. \href{https://drive.google.com/file/d/16QsLSrpR1ejFFz4tK7IcxxIqv8V7Z55y/view?usp=share_link}{\color{blue}Movie here}
    \item S4: This is a simulation movie for $\ell=2.5$ mm, $\phi=0.85$. \href{https://drive.google.com/file/d/13NVRrv0LM8qyO8e28VXAfEBswUE9F7H5/view?usp=share_link}{\color{blue}Movie here}
    \item S5: This is a simulation movie for large rod length and high area fraction with $\ell=4.5$ mm, $\phi=0.85$ when the system forms is in the layered phase. \href{https://drive.google.com/file/d/1wDHbm2wj9UXAuMC0eYGRXU9VvbeOhUlE/view?usp=share_link}{\color{blue}Movie here}
    \item S6: Trajectories of single rod at $\phi=0.65$ and $\phi=0.85$ for $\ell=2.5$ mm. \href{https://drive.google.com/file/d/1K3jc_GQElyyMdMnS7_VZRMgL9oVOXOeg/view?usp=share_link}{\color{blue}Movie here}
    \item S7: This is a simulation movie for $\ell=3.0$ mm, $\phi=0.85$; the motile defects are clearly visible. \href{https://drive.google.com/file/d/1uUD35ypdpqA0xoWv6U_OhaFCWnW0jydu/view?usp=share_link}{\color{blue}Movie here}
\end{itemize}

\subsection*{Steady state configuration of $\ell=3.5$ mm, $\phi=0.85$}

\begin{figure}[h!]
    \centering
    \includegraphics[width=0.47\textwidth,height=0.47\textwidth]{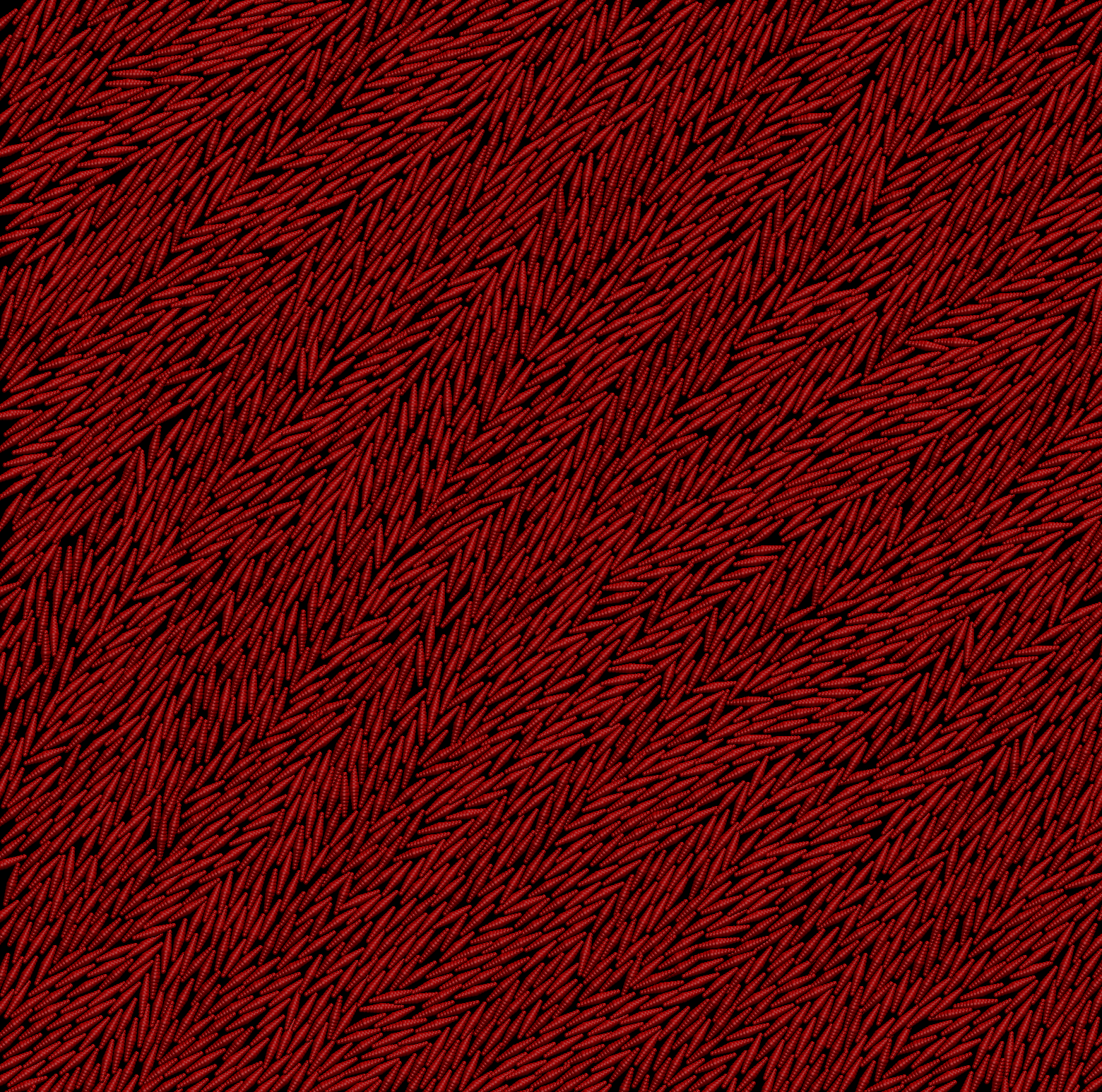}
    \caption{Steady state configuration of the rods for $\ell=3.5$ mm at $\phi=0.85$;  rods are aligned in a direction other than $x$ or $y$ axes.}
    \label{l35f85}
\end{figure}

\subsection*{$G_2(\mathbf{r})$ vs $\mathbf{r}$ for different gap values in base and lid ($w$)}
\begin{figure}[h!]
 \input{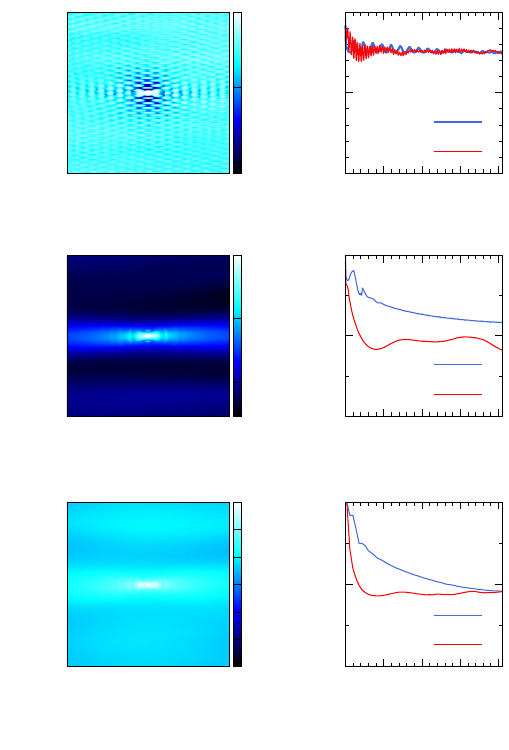}
    \caption{Nematic order parameter correlation function $G_2(\mathbf{r})$ for $\ell =4.5$ mm, $\phi=0.80$ for different $w$ values. (a) Heat map and (b) $G_2(\mathbf{r})$ vs $r$  parallel and perpendicular to alignment for $w = 0.90$ mm. (c) Heat map and (d) $G_2(\mathbf{r})$ vs $r$ parallel and perpendicular to alignment for $w=1.4$ mm . (e) Heat map and (f) $G_2(\mathbf{r})$ vs $r$ along and normal to the alignment direction for $w=1.6$.}
    \label{g2r}
\end{figure}
\end{widetext}

\end{document}